\documentclass[reprint,prd,nofootinbib,superscriptaddress]{revtex4}
\usepackage{CJK}
\usepackage{amsfonts}
\usepackage{amsmath,slashed}
\usepackage{amssymb}
\usepackage[english]{babel}
\usepackage{graphicx}
\usepackage{epsfig}
\usepackage{bm}
\usepackage{verbatim}
\usepackage[utf8]{inputenc}
\usepackage{booktabs}
\usepackage{multirow}
\usepackage{subfig}
\usepackage{slashed}
\usepackage{xcolor}
\usepackage[colorlinks=true,urlcolor=red,citecolor=red]{hyperref}
\usepackage[font=small]{caption}
\usepackage{float}
\usepackage{blindtext}
\usepackage{placeins}

\newcommand{\hs}{\hspace*{0.3cm}}

\newcommand{\vs}{\vspace*{0.5cm}}
\newcommand{\be}{\begin{equation}}
\newcommand{\ee}{\end{equation}}
\newcommand{\bea}{\begin{eqnarray}}
\newcommand{\eea}{\end{eqnarray}}
\newcommand{\ben}{\begin{enumerate}}
\newcommand{\een}{\end{enumerate}}
\newcommand{\bit}{\begin{itemize}}
\newcommand{\eit}{\end{itemize}}
\newcommand{\bde}{\begin{widetext}}
\newcommand{\ede}{\end{widetext}}
\newcommand{\bib}{\bibitem}
\newcommand{\nn}{\nonumber}
\newcommand{\crn}{\nonumber \\}

\newcommand{\al}{\alpha}
\newcommand{\la}{\lambda}
\newcommand{\bet}{\beta}
\newcommand{\ga}{\gamma}
\newcommand{\va}{\varphi}
\newcommand{\om}{\omega}
\newcommand{\pa}{\partial}

\newcommand{\fr}{\frac}

\newcommand{\bc}{\begin{center}}
\newcommand{\ec}{\end{center}}

\newcommand{\de}{\delta}
\newcommand{\De}{\Delta}
\newcommand{\ep}{\epsilon}

\newcommand{\La}{\Lambda}
\newcommand{\si}{\sigma}
\newcommand{\Si}{\Sigma}

\newcommand{\Om}{\Omega}
\newcommand{\eq}{\eqref}
\newcommand{\lb}{\label}

\newcommand{\mathsym}[1]{{}}
\newcommand{\gev}{~\mathrm{GeV}}
\newcommand{\tev}{~\mathrm{TeV}}

\definecolor{bostonuniversityred}{rgb}{0.8, 0.0, 0.0}

\def\gsim{\raise0.3ex\hbox{$\;>$\kern-0.75em\raise-1.1ex\hbox{$\sim\;$}}}
\def\lsim{\raise0.3ex\hbox{$\;<$\kern-0.75em\raise-1.1ex\hbox{$\sim\;$}}}

\begin{document}
	\title{New Physics  in the 3-3-1 models}
	\author{H. N. Long}
	\email{hoangngoclong@vlu.edu.vn}
	\affiliation{Subatomic Physics Research Group,
		Science and Technology Advanced Institute,\\
		Van Lang University, Ho Chi Minh City, Vietnam\\
	and\\
 Faculty of Applied Technology, School of  Technology,  Van Lang University, Ho Chi Minh City, Vietnam}

\date{\today }

\begin{abstract}
Two main ingredients of current particle physics 
 such as local gauge symmetry and mass generation via the Higgs mechanism being  basic ground of the Standard Model are widely confirmed by 
  experimental data.  However, some problems   such as neutrino masses, 
dark matter, baryon asymmetry of Universe have clearly indicated  that the Standard Model cannot be the ultimate theory of nature. To surpass the mentioned puzzles, many extensions of  the Standard Model (called beyond  Standard Model) have been proposed. Among  beyond  Standard Models, the 3-3-1 models have some intriguing features and they get wide attention. The pioneer models develop in some directions.  In this paper, 
new main  versions   of the  3-3-1 models and their  consequences  are presented.

\end{abstract}


\maketitle
\noindent

\section{Introduction}
\lb{sec1} 
The Standard Model (SM) has been very successful in describing observed phenomena. However, it also leaves many striking features of the physics of our world unanswered ~~\cite{pdg}. First of all, electric charge and family number in the standard model are in principle arbitrary, opposite to observation. Further, neutrino oscillations coming from the sun, the atmosphere, reactors and accelerators have been well-confirmed, implying that neutrinos have nonzero masses and mixing, in contrast to the SM   that possesses massless neutrinos. Alternatively, dark matter (DM) that makes up most of the mass of galaxies and galactic clusters is completely not present in the SM. 
 Last, but not least, the SM
  cannot explain the baryon-number asymmetry of the universe, for which we nowadays observe only matter, and there is no popular existence of antimatter.            

That said, it is well-established that the SM 
of elementary particles and interactions has to be extended. More specifically, the most problems arise in the electroweak sector ~\cite{criv}. Among beyond Standard Models (BSMs), the models based on the  $\mbox{SU}(3)_C\otimes \mbox{SU}(3)_L \otimes \mbox{U}(1)_X$ gauge group (called by 3-3-1 models) ~\cite{ppf1,ppf2,ppf3,ppf4,flt1,flt2,flt3,flt4,flt5,flt6} have some intriguing features such as the models can give explanation for generation number ($n_f$ to be three)  due to discrimination in the quark sector  (one  quark family
transforms differently from other twos) leading to  ($n_f = 3\times n_C$) 
and asymptotic freedom in QCD ($n_f \leq 5$). The electric charge quantization in the 3-3-1 models is related to their fermion representation structures under 3-3-1 symmetry. Another interesting  feature should be mentioned concerning automatic satisfaction with Peccei-Quinn mechanism ~\cite{pal}.  However,  the 3-3-1 models contain one disadvantage: due to discrimination of quark generations, the scalar sector is quite complicate. There are attempts to overpass the problem.

The 3-3-1 models presently attract much attention concerning their ability to address the questions of neutrino masses and 
DM. Besides, they present potential signals at colliders, associated with the questions of FCNCs. It is stated that the 3-3-1 model is the first non-abelian gauge principle that recognizes dark matter stability. Additionally, compelling neutrino mass generation mechanisms, such as canonical seesaw, type II seesaw, inverse seesaw, and scotogenic scheme, are all realized in the 3-3-1 models.      

The aim of this work is to summarize new developments and some interesting consequences in the 3-3-1 models.
The rest of this article is organized as follows. The section \ref{sec2} is devoted to pioneer 3-3-1 models and their minimal scalar contents. In section \ref{sec3} briefly reviews the model of discrimination of leptons and
model with inverse seesaw.  Combination of non-Abelian discrete symmetries with the  3-3-1 models is presented in section \ref{sec4}. Section \ref{sec5} is devoted to  newest development in direction of axion or axion-like particle (ALP).  In section \ref{s6}, I just mention the main phenomenologies of the models such as collider physics, 
neutrino mass, LFV, $g-2$ and Early Universe. The conclusions are in the last section \ref{sec7}.
 
\section{The pioneer  3-3-1  models }
\lb{sec2}
In the pioneer models, the difference  mostly is   
in content of the lepton triplet: In the minimal 3-3-1 model (M331 model)
~\cite{ppf1,ppf2,ppf3,ppf4}, the
right-handed charged lepton is at the bottom of the  lepton triplet, while in the 3-3-1 model with right-handed neutrinos (331RN  model) ~\cite{flt1,flt2,flt3,flt4,flt5,flt6}, right-handed neutrino is at the bottom of the lepton triplet instead. 
 
  The electric charge operator is expressed in the  form of diagonalized ones:
 \be Q= T_3 + \bet T_8  + X, \label{ctdt1} \ee
where the factor one  ($1$) associated with $T_3$ ensures  embedding in  of the SM. It is worth mentioning that   the 3-3-1 models are characterized by the value of $\bet$. 
 
\subsection{The minimal 3- 3-1 model}
\lb{sec21}
 Leptons  come in fundamental representatin of $\mbox{SU}(3)_L $  ($\bet = -\sqrt{3}$)~\cite{ppf3}:
 \be f_{a  L}^T = \left( \nu_{a L}, 
 l_{a L},    (l^c)_{a L}, \right) \sim (1, 3 , 0),
\label{l} \ee  where  $ a = 1, 2, 3$ is generation index. Note that if {\it leptons lie
in antitriplet, then the value of $\bet$ will just change the sign}, namely, in this case  
$\bet = +\sqrt{3}$).

The third   quark generation is in triplet and two others  are  in
anttriplet:
  \bea
 Q_{3L}^T & = & \left(
u_{3L},    d_{3L},    T_{L} \right) \sim (3, 3, 2/3),\crn 
  Q_{iL}^T & =  &\left( d_{iL}, - u_{iL},   D_{iL}
 \right) \sim (3, \tilde{3}, - 1/3 ), \label{q} \\
 u_{iR} &\sim & (3, 1, 2/3), d_{iR}\sim (3, 1, -1/3),
D_{iR}\sim (3, 1, -4/3),\ i=1,2,\crn
 u_{3R}&\sim & (3, 1, 2/3), d_{3R}\sim (3, 1, -1/3), T_{R}
\sim (3, 1, 5/3).\nn\eea

The  spontaneous symmetry
breakdown  (SSB)    of the model requires 
three scalar triplets and one sextet 
 \bea \chi^T & = &\left(  \chi_1^{-} ,  \chi_2^{--} ,  \chi_3^{0}
  \right) \sim (1, 3, - 1),\label{mh1}\\
 \rho^T & =& \left( 
   \rho^+_1\ , \rho_2^0\ , \rho^{++}_3
  \right) \sim (1, 3, 1),\crn
 \eta^T & =& \left( \eta_1^{0},  \eta_2^{-},  \eta_3^{+}
  \right) \sim (1, 3, 0),\crn
S & = & \left( \begin{array}{ccc}
 \si_1^0 &h_2^{-}/ \sqrt{2} &  h_1^+/ \sqrt{2}\\
h_2^{-}/ \sqrt{2} & H_1^{--} & \si^0_2/ \sqrt{2}\\
 h_1^+/ \sqrt{2} & \si^0_2/ \sqrt{2} & H_2^{++}
 \end{array}  \right) \sim (1, 6, 0).\nn\eea
  with VEVs:
 \bea \langle \chi \rangle^T  & = &\left( 0 , 0 ,  \om/\sqrt{2}
  \right) \,,\,
\langle \rho \rangle^T  = \left( 
   0\ , v /\sqrt{2}\ , 0
  \right)
  \,, \,
 \langle \eta \rangle^T  = \left( u /\sqrt{2},  0,  0
\right) ,\crn
 \sqrt{2} \, \langle S \rangle & = & \left( \begin{array}{ccc}
0 & 0 &  0\\
0 & 0 & v^\prime \\
 0 & v^\prime  & 0
 \end{array}  \right)\,.
 \lb{pbmin}
 \eea 
Without  right-handed charged lepton, the lepton mass matrix is antisymmetric $3 \times 3$ matrix
containing one massless eigenvalue. To solve this puzzle, the sextet with symmetric interaction to leptonic triplets, has been added ~\cite{ppf3}.  

The SSB    of the model is in two steps as follows:
\be \mathrm{SU}(3)_L
\otimes \mathrm{U}(1)_X \stackrel{ \langle \chi_3^0\rangle }{\longrightarrow}\ \mathrm{SU}(2)_L \otimes
\mathrm{U}(1)_Y\stackrel{ u, v , v^\prime }{\longrightarrow} \mathrm{U}(1)_{Q},\label{st1}\ee  
The ratio between couplings constant of $U(1)_X$ and $SU(3)_L$ is given by ~~~\cite{ppf3}
\be
t = \fr{g_X} g = \fr{\sin_W^2}{1 - 4 \sin_W^2}\,.
\lb{valuet}
\ee
Denominator in Eq. \eq{valuet} tends to zero  (Landau pole) ~\cite{ppf4,landau1} at 5 TeV. In recent work, 
by adding scalar leptoquarks, this  pole has been searched at 100 TeV ~\cite{landau2}. 

\subsection{The 3-3-1  model with  right-handed neutrinos}
\label{sec22}
 In the    331RN  model, leptons are in  triplet ~~\cite{flt2,flt3,flt4,flt5,flt6} ($\bet = -\fr 1{ \sqrt{3}}$):
  \be f_{a L}^T= \left(
  \nu_{a L },  l_{a L },  (\nu^c_L)_{a  }
  \right) \sim (1, 3, -1/3), l^a_R\sim (1, 1, -1), \label{l21} \ee
where  $ a = 1, 2, 3$ is a generation index. Two first generation
of quarks comes  in the anti-fundamental representation of $\mbox{SU}(3)_L $, and the third one is in triplet:
\be Q_{iL}^T = \left(  d_{iL}, -u_{iL},  D_{iL}
 \right) \sim (3, \tilde{3}, 0), \label{q2} \ee
\[ u_{iR}\sim (3, 1, 2/3), d_{iR}\sim (3, 1, -1/3),
D_{iR}\sim (3, 1, -1/3),\ i=1,2,\] \[
 Q_{3L}^T = \left(  u_{3L},  d_{3L},  T_{L}
 \right)^T \sim (3, 3, 1/3),\]
\[ u_{3R}\sim (3, 1, 2/3), d_{3R}\sim (3, 1, -1/3), T_{R}
\sim (3, 1, 2/3).\]
 To spontaneous symmetry breaking, we need three  Higgs
triplets
 \bea \chi^T & = &\left( \chi^0, \chi^-,  \chi^{\prime 0}
 \right) \sim (1, 3, -1),\crn
 \rho^T & =& \left( \rho^+,  \rho^0,  \rho^{\prime +}
 \right) \sim (1, 3, 2),\label{h1}\\
\eta^T & =& \left(  \eta^0, \eta^-,  \eta^{\prime 0}
 \right) \sim (1, 3, -1).\nn \eea
 The SSB follows the scheme as below \be \mathrm{SU}(3)_L
\otimes \mathrm{U}(1)_X \stackrel{ \langle \chi^{\prime 0}
\rangle}{\longrightarrow}\ \mathrm{SU}(2)_L \otimes
\mathrm{U}(1)_Y\stackrel{ \langle \rho \rangle,\langle \eta
\rangle}{\longrightarrow} \mathrm{U}(1)_{Q},\label{st1}\ee  where
 \bea \sqrt{2}\,  \langle \chi \rangle   = \left( 0 , 0 ,  \om
  \right)^T \,,\hs 
\sqrt{2}\, \langle \rho \rangle  = \left( 
   0\ , v \ , 0
  \right)^T\, ,\hs 
\sqrt{2}\,  \langle \eta \rangle  = \left( u ,  0,  0
  \right)^T .\lb{pvrhn}
 \eea 
Note that the model in which $\bet = 0$ has been constructed in ~\cite{hueninh1} and its
phenomenology in quark sector  was studied in ~\cite{hueninh2}. If so, the model contains extra lepton $E_a$
with electric charge equal to $ -1/2$. 

The models with arbitrary $\bet$ (331$\bet$) were proposed in Ref. ~\cite{331bet}. 

As above mentioned, the 3-3-1 models contain many intriguing features, but they face one limitation in
large scalar content which  prevent  
  their  predictability.
 A reason for this is that  due to discrimination in quark/lepton representations,
more scalar multiplets are required. 
There are attempts in this direction and results are two models: economical (E331) and simple (S331) models.
  
\subsection{The economical  3-3-1 model}
\lb{sec23}
Note that in the model with $\bet =- \fr 1{\sqrt{3}}$, the exist two triplets $\eta$ and $\chi$ 
 with identical quantum.  Hence, we can omit one
$\eta$~\cite{e331p,e3311,dlahep}, then 
 \bea \chi^T & = &\left(
 \chi^0,  \chi^-,  \chi^{\prime \, 0}  \right) \sim (1, 3, -1),
\label{h1}\\
 \rho^T & =& \left( \rho^+,  \rho^0,  \rho^{\prime +}
 \right) \sim (1, 3, 2)\,. \label{betas1} \eea \hs To provide masses of fermions and gauge
 bosons,  
 one needs to provide  $\chi$ with the following  VEV: \be
\sqrt{2}\, \langle \chi \rangle^T = ( u ,\ 0, \om )\,.\label{betas1t} \ee
It is worth mentioning that the triplet $\chi$ contains the VEV $\om$ which responds for the first step of SSB. 

  One of the VEVs $u$ carrying lepton number two  is a source of lepton-number
violations and a reason for the  mixing between $W$ and  singly-charged bilepton
gauge boson  $Y$ as well as between  $X^0$ and
the photon, the $Z$ and the $Z^\prime$.
Thus, this leads to $u
\ll v$, and there are three quite different scales for the VEVs of
the model: one is very small $u \simeq \emph{O}(1) \gev $  - a
lepton-number violating parameter, the second $v$ is close to the
SM one : $v\simeq v_{weak} = 246 \gev$ and the last is in the range
of new physics scale about $\emph{O}(1) \tev$.

Due to $W-Y$ mixing, there exist lepton number violating (LNV)  couplings
in both $W$ and $Y$ interactions 
  as follows  ~~\cite{e3311}
\be \sqrt{2}\, H^{\mathrm{CC}}= g\, \left(j^{\mu-}_W W^+_\mu +
j^{\mu-}_Y Y^+_\mu + j^{\mu 0*}_X X^{0}_\mu + H.c.\right)
 \ee
where \bea j^{\mu-}_W&=&c_\theta (\overline{\nu}_{iL}\ga^\mu
e_{iL}+\overline{u}_{iL}\ga^\mu d_{iL})+s_\theta
(\overline{\nu}^c_{iL}\ga^\mu
e_{iL}+\overline{U}_{L}\ga^\mu d_{1L}+\overline{u}_{\al L}\ga^\mu D_{\al L}),\label{dgw}\\
j^{\mu-}_Y&=&c_\theta (\overline{\nu}^c_{iL}\ga^\mu
e_{iL}+\overline{U}_{L}\ga^\mu d_{1L}+\overline{u}_{\al L}\ga^\mu
D_{\al L})-s_\theta (\overline{\nu}_{iL} \ga^\mu
e_{iL}+\overline{u}_{i L}\ga^\mu d_{i L}),\label{dgy}\\
j^{\mu 0*}_X &=& (1-t^2_{2\theta})(\overline{\nu}_{iL}\ga^\mu
\nu^c_{i L}+\overline{u}_{1L}\ga^\mu U_{L}-\overline{D}_{\al
L}\ga^\mu d_{\al
L})-t^2_{2\theta}(\overline{\nu}^c_{iL}\ga^\mu\nu_{iL}+\overline{U}_L\ga^\mu
u_{1L}-\overline{d}_{\al L}\ga^\mu D_{\al L})\crn
&&+\fr{t_{2\theta}}{\sqrt{1+4t^2_{2\theta}}}(\overline{\nu}_i\ga^\mu
\nu_i+\overline{u}_{1L}\ga^\mu u_{1L}-\overline{U}_L\ga^\mu
U_L-\overline{d}_{\al L}\ga^\mu d_{\al L}+\overline{D}_{\al
L}\ga^\mu D_{\al L}).\label{dgx} \eea
Note that the LNV couplings lie in the second term of Eq. \eq{dgw} and in the first term of Eq. \eq{dgy}.

\begin{figure}[h]
 	\centering
 	\includegraphics[width=0.5\textwidth]{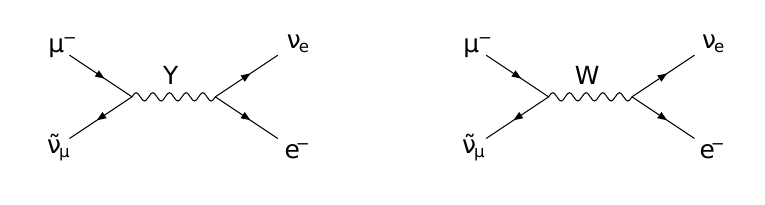}
 	\caption{\label{ht} Feynman diagrams for wrong muon decay ($\mu^- \to e^- \nu_e \tilde{\nu}_\mu$). The new contribution is presented in the second diagram (only in the economical 3-3-1 model)}.
\end{figure}

Some interesting phenomenologies can be illustrated by figure ~\ref{ht} bellows 

It is emphasized that the first diagram  exists in both pioneer and E331 models, while the second diagram exists  only in the E331 model.

Due to  the minimal scalar sector, some quarks  have vanishing masses at the tree-level,
 and they  get masses  at loop level ~~~\cite{e331qm1,e331qm2,e331qm3}. 

The model is very rich in neutrino physics, and it has only one problem: no candidate for dark matter.
For this purpose, the inner  triplets were added ~~\cite{e331n}. 

\subsection{The simple  3-3-1 model}
\lb{sec24}
This model belongs to class with $\bet = - \sqrt{3}$ and the  third generation of quarks  transforms differently from the first two generations ~\cite{S331}:
\bea \psi_{aL}^T &\equiv & \left(
               \nu_{aL}  \, ,  l_{aL} \,,  (l_{aR})^c
\right) \sim (1,3,0),\, 
Q_{\al L}^T  \equiv \left(
  d_{\al L}\,,  -u_{\al L}\,,  D_{\al L}
\right) \sim (3,\tilde{3} ,-1/3),\crn
 Q_{3L}^T &\equiv& \left( u_{3L}\,,  d_{3L}\,,  T_{L} \right) \sim
 \left(3,3,2/3\right), \\ u_{a
R}&\sim&\left(3,1,2/3\right),\hs d_{a R} \sim \left(3,1,-1/3\right),\, 
D_{\al R}  \sim
\left(3,1,-4/3\right),\hs T_{R} \sim \left(3,1,5/3\right),\nn \eea where $a=1,2,3$
and $\al= 1,2$ are family indices.  The new quarks possess exotic electric charges as $Q(D_\al)=-4/3$ and $Q(T)=5/3$.

The model has  only two scalar triplets ~\cite{S331,r331}:
 \bea \eta^T = \left(
\eta^0_1\,, 
\eta^-_2\,, 
\eta^{+}_3 \right) \sim (1,3,0),\hs \chi^T = \left(
\chi^-_1\,, 
\chi^{--}_2\,, 
\chi^0_3 \right) \sim (1,3,-1),\label{vev2}
\eea with VEVs \be  \sqrt{2}\, \langle \eta\rangle^T = 
\left(
u\,,  0\,,  0 \right) ,\hs  \sqrt{2}\, \langle \chi\rangle^T = 
\left(
0\,,  0\,,  \om \right).\ee  
To have DM candidate, sextet with $X=1$ was added   ~\cite{S331}:
\be S=\left(\begin{array}{ccc}
h^+_{11} & \fr{\si^{0}_{12}}{\sqrt{2}} & \fr{h^{++}_{13}}{\sqrt{2}}\\ 
\fr{\sigma^{0}_{12}}{\sqrt{2}} & h^{-}_{22} & \fr{h^+_{23}}{\sqrt{2}} \\
\fr{h^{++}_{13}}{\sqrt{2}} & \fr{h^+_{23}}{\sqrt{2}} & h^{+++}_{33}
\end{array}
\right)\sim (1,6,1).\ee

Under a $Z_2$ symmetry,  this sextet is  odd , whereas all the other fields are even.

Because of the minimal scalar sector, some fermions are massless at the tree-level. However, they can get corrections coming from the effective interactions.

\section{The  3-3-1 model specific for leptons and neutrinos}
\lb{sec3}
According LHCb data in 2014, the original  3-3-1 models face  non-universality in lepton sector  and neutrino physics. The attempts for solving these troubles
 were  made and some of them are presented below.

\subsection{The flipped 3-3-1 model}
\lb{sec31}
In the original 3-3-1 models, the anomaly free requires one quark generation transforms under $SU(3)_L$  differently from  two the other ones. In the flipped version, the discrimination happens in lepton sector 
~~\cite{f3311}, namely, the lightest leptons ($e$) transforms as sextet (symmetric representation), while two heavy ($\mu$) and ($\tau$) transform as triplets. In contrast, all left-handed quarks are in antitriplets, so that the model is free gauge anomaly. To provide fermion masses, the scalar sector contains three
triplets and one sextet. The particle content of the model is summarized  in Table \ref{fmolel1}
~\cite{f3311,f3312}

\begin{table}[ht]
	\centering
		\begin{tabular}{c|c|c|c|c}	
			Name & 3-3-1 rep.
			& Components & \# flavors
			\tabularnewline
			  \hline
			$L_{e}$ & $\left(\mathbf{1},\mathbf{6},-\fr{1}{3}\right)$ &
			$\left(\begin{array}{ccc}
			\left( \Si^{-}_R\right)^c & \fr{1}{\sqrt{2}}\Si^{0}_L & \fr{1}{\sqrt{2}}\nu_{eL}\\
			\fr{1}{\sqrt{2}}\Si^{0}_L & \Si^{-}_L & \fr{1}{\sqrt{2}}l_L\\
			\fr{1}{\sqrt{2}}\nu_{lL} & \fr{1}{\sqrt{2}}l_L & E_{lL}
			\end{array}\right)$ & 1
			\tabularnewline
			 \hline
			$L_{\al=\mu,\tau}$ 
			& $\left(\mathbf{1},\widehat{\mathbf{2}},-\fr{1}{2}\right)+\left(\mathbf{1},\widehat{\mathbf{1}},-1\right)$ & $\left(\nu_{\al},l_{\alpha},E_{\alpha}\right)^{T}_L$ & 2
			\tabularnewline
			 \hline
			$l_{\al {R}}$ 
			 & $\left(\mathbf{1},\widehat{\mathbf{1}},-1\right)$ & $l_{\alpha{R}}$ & 6\tabularnewline
			   \hline
			$Q_{\alpha}$ 
			& $\left(\mathbf{3},\widehat{\mathbf{2}},\fr{1}{6}\right)+\left(\mathbf{3},\widehat{\mathbf{1}},\fr{2}{3}\right)$ & $\left(d_{\al},-u_{\al},U_{\al}\right)^{T}_{L}$ & 3\tabularnewline
			 \hline
			$u_{\al{R}}$ 
			 & $\left(\mathbf{3},\widehat{\mathbf{1}}, \fr{2}{3}\right)$ & $u_{\al {R}}$ & 6\tabularnewline
			  \hline
			$d_{\al{R}}$ 
			& $\left( \mathbf{3},\widehat{\mathbf{1}}, -\fr{1}{3}\right)$ & $d_{\alpha{R}}$ & 3\tabularnewline
			 \hline
			$\phi_{i=1,2}$ 
			& $\left(\mathbf{1},\widehat{\mathbf{2}},\fr{1}{2}\right)+\left(\mathbf{1},\widehat{\mathbf{1}},0\right)$ & $\left(H_{i}^{+},H_{i}^{0},\si_{i}^{0}\right)^{T}$ & 2\tabularnewline
			 \hline
			$\phi_{3}$ 
			& $\left(\mathbf{1},\widehat{\mathbf{2}},-\fr{1}{2}\right)+\left(\mathbf{1},\widehat{\mathbf{1}},-1\right)$ & $\left(H_{3}^{0},H_{3}^{-},\sigma_{3}^{-}\right)^{T}$ & 1\tabularnewline
			 \hline
			$S$ 
			& $\left(\mathbf{1},\widehat{\mathbf{3}},1\right)+\left(\mathbf{1},\widehat{\mathbf{2}},\fr{1}{2}\right)+\left(\mathbf{1},\widehat{\mathbf{1}},0\right)$ & $\left(\begin{array}{ccc}
			\De^{++} & \fr{1}{\sqrt{2}}\De^{+} & \fr{1}{\sqrt{2}}H_{S}^{+}\\
			\fr{1}{\sqrt{2}}\De^{+} & \De^{0} & \fr{1}{\sqrt{2}}H_{S}^{0}\\
			\fr{1}{\sqrt{2}}H_{S}^{+} & \fr{1}{\sqrt{2}}H_{S}^{0} & \si_{S}^{0}
			\end{array}\right)$ & 1\tabularnewline
  \hline
		\end{tabular}\vs
		
		\caption{\lb{fmolel1}Particle content of  the flipped 3-3-1 model.}
\end{table}

The  Lagrangian for  leptons is given as
\be \label{yufm}
-\mathcal{L}^Y_{\mathrm{lepton}}=\sum_{i=1}^2 \sum_{\al=\mu,\tau}\sum_{\beta=1}^6y^{\ell(i)}_{\al\beta}\overline{l_{\beta{R}}} L_{\al}\phi^*_i +\sum_{\beta=1}^6y^{\ell\prime}_{\beta} \overline{e_{\beta{R}}}L_l S^* +y^{\ell\prime\prime} \overline{(L_l)^c}L_lS +\mathrm{H.c.},
\ee 
where the  invariant term  of the tensor product of the three sextets is expanded as  $\overline{(L_l)^c}L_lS=\ep^{abc}\ep^{ijk}\overline{(L_l)^c}_{ai}(L_l)_{bj}S_{ck}$, 
 $(L_l)^c_{ai}\equiv C\overline{(L_l)_{ai}}^T$.  Note that $\phi_3$  only appears in the Yukawa part of the quark  guaranteeing that all quark are massive. In contrast, at tree level, the electron and neutrinos are massless, but one-loop contributions are enough to.

Note that there is no  generation discrimination between three left-handed quarks multiplets  in the flipped 3-3-1 model. 
Hence, there  does not exist flavor changing neutral currents mediated by extra $Z^\prime$ at tree level. The natural mixing happening in the lepton sector of the flipped 3-3-1 predicts  many lepton flavor violating (LFV) sources. It leads  to the existence of the LFV decays, which  should be investigated thoroughly elsewhere. For example the LFV decays of the SM-like Higgs boson was discussed in Ref. ~~\cite{f3312}. Also, the new heavy leptons in the electron sextet may gives significant one-loop contributions to $(g-2)_e$ anomalies. 

Note that, in the flipped 3-3-1 model, due to lepton generation discrimination, there exists lepton flavor  changing neutral current mediated by extra $Z^\prime$ at tree level. 

\subsection{The  3-3-1 model  with inverse seesaw neutrinos}
\lb{sec32}
It is well known that the seesaw mechanism happens at very high energy scale. The inverse seesaw (ISS) mechanism
is one of the mechanisms with low energy scale,  which can explain the neutrino oscillation as well as the  lepton flavor violating decay rates of charged leptons (cLFV) hunted by experiments.
The  3-3-1 model  with ISS neutrinos (331ISS) for  $\bet = \fr 1{\sqrt{3}}$ has been constructed in  Refs. ~\cite{Dias:2012xp, Catano:2012kw, in331,in332}.
 The quark sector 
  is  similar and quark discussion is  referred to Ref~.\cite{in331}. Each  lepton family consists of  a  $SU(3)_L$ triplet
  \[\psi^T_{aL}= (\nu_a,~l_a, N_a)_L\sim \left(1, 3,-\fr{1}{3}\right)\,,\]
   and a right-handed charged lepton  $l_{aR}\sim (1,1,-1)$ with $a=1,2,3$.  Each left-handed neutrino $N_{aL}=(N_{aR})^c$  is equivalent with  a new right-handed neutrinos defined in previous  331RN  model~\cite{flt2}.  
 The 331ISS model contains  three more RH neutrinos transforming as gauge singlets, $X_{aR}\sim (1, 1,0)$, $a=1,2,3$. They interact with  
 the $SU(3)_L$ Higgs triplets to generate the neutrino mass term relating with the ISS mechanism. 
 
 The  Higgs sector is the same as  in Section \ref{sec22}.
 
The  Lagrangian for generating lepton masses is ~\cite{in333}:
\begin{align}
\label{yukj172}
\mathcal{L}^{\mathrm{Y}}_{lep} =-h^e_{ab}\overline{\psi_{aL}}\rho l_{bR}+
h^{\nu}_{ab} \epsilon^{ijk} \overline{(\psi_{aL})_i}(\psi_{bL})^c_j\rho^*_k
- Y^{ab}\overline{\psi_{aL}}\,\chi X_{bR} -\fr{1}{2} (\mu_{X})^*_{ba}\overline{(X_{aR})^c}X_{bR}+ \mathrm{H.c.}.
\end{align}
Assuming  the model  respects a new lepton number symmetry $\mathcal{L}$   discussed  in  ~\cite{longchang}  so that   the term   $\overline{\psi_{aL}}\,\eta X_{bR}$ is not allowed in the above Yukawa Lagrangian, while the soft-breaking term $(\mu_{X})^*_{ba}\overline{(X_{aR})^c}X_{bR}$ is allowed with small $(\mu_{X})_{ba}$. The new lepton number $\mathcal{L}$ called by generalized lepton number ~~~\cite{jhep18} is defined as $L=\fr{4}{\sqrt{3}} T^8 + \mathcal{L} \mathbb{I}$, where $L$ is  the normal lepton number.  

The first term in Lagrangian \eq{yukj172} generates charged lepton masses $m_{e_a}\equiv \fr{h^e_{ab}v_1}{\sqrt{2}} \de_{ab}$,
 i.e, the mass matrix of the charged leptons is assumed to be diagonal,  hence  the flavor states of the charged leptons are also the physical ones. In the basis $\nu^{\prime T}_{L}=(\nu_{L}, N_{L}, (X_R)^c)$ and $(\nu^\prime _{L})^c=((\nu_L)^c, (N_L)^c, X_R)^T$ of the neutral leptons,   Lagrangian \eqref{yukj172} gives a neutrino mass term corresponding to a  block form of the mass matrix~\cite{in332}, namely
\begin{align}
-{\mathcal{L}}^{\nu}_{\mathrm{mass}}=\fr{1}{2}\overline{\nu'_L}M^{\nu\dagger }(\nu'_L)^c +\mathrm{H.c.}, \,\mathrm{ where }\quad M^{\nu\dagger}=\begin{pmatrix}
0	& m_D &0 \\
m_D^T	&0  & M_R \\
0& M_R^T& \mu_X^{\dagger}
\end{pmatrix},  \label{Lnu1}
\end{align}
where   $M_R$  is  a $3\times3$ matrix  $(M_R)_{ab}\equiv Y_{ab}\fr{w}{\sqrt{2}}$, $(m_D)_{ab}\equiv \sqrt{2}h^{\nu}_{ab}v_1$ with $a,b=1,2,3$. Neutrino subbases are denoted as  $\nu_{R}^T=((\nu_{1L})^c,(\nu_{2L})^c,(\nu_{3L})^c)$, $N_R^T=((N_{1L})^c,(N_{2L})^c,(N_{3L})^c)$, and  $X_L^T=((X_{1R})^c,(X_{2R})^c,(X_{3R})^c)$.
 The mass matrix $M_R$ does not appear in the 331RN  model.  The Dirac neutrino mass matrix $m_D$ must be antisymmetric.  The matrix  $\mu_{X}$ defined in Eq.~\eqref{yukj172}  is symmetric and it can be diagonalized by a transformation $U_X$:
\be\label{28t}
	U_X^T\mu_{X}U_X=\mathrm{diag}\left(\mu_{X,1},\mu_{X,2},\mu_{X,3}\right).
\ee
The matrix $U_X$ will be absorbed by redefinition of the  states $X_{a}$, therefore $\mu_X$ will be set as the diagonal matrix given in the right-hand side of  Eq.~\eqref{28t}.

The mass matrix  $M^{\nu}$ is   diagonalized by a $9\times9$ unitary matrix $U^{\nu}$,
\begin{align}
U^{\nu T}M^{\nu}U^{\nu}=\hat{M}^{\nu}=\mathrm{diag}(m_{n_1},m_{n_2},..., m_{n_{9}})=\mathrm{diag}(\hat{m}_{\nu}, \hat{M}_N), \label{pt28m}
\end{align}
where $m_{n_i}$ ($i=1,2,...,9$) are  masses of the nine physical neutrino states $n_{iL}$. They consist of  three active neutrinos  $n_{aL}$ ($a=1,2,3$) corresponding to the mass  submatrix    $\hat{m}_{\nu}=\mathrm{diag}(m_{n_1},\;m_{n_2},\;m_{n_3})$, and the six extra neutrinos $n_{IL}$ ($I=4,5,..,9$)  with  $\hat{M}_N=\mathrm{diag}(m_{n_4},\;m_{n_5},...,\;m_{n_{9}})$.
The ISS mechanism leads to the following approximate solution of $U^{\nu}$,
\begin{align}
\label{pt29t}
U^{\nu}= \Om \left(
\begin{array}{cc}
U_{\mathrm{PMNS}} & \mathbf{O} \\
\mathbf{O} & V \\
\end{array}
\right), \;\; \Om=\exp\left(
\begin{array}{cc}
\mathbf{O} & R \\
-R^\dagger & \mathbf{O} \\
\end{array}
\right)=
\left(
\begin{array}{cc}
1-\fr{1}{2}RR^{\dagger} & R \\
-R^\dagger &  1-\fr{1}{2}R^{\dagger} R\\
\end{array}
\right)+ \mathcal{O}(R^3),
\end{align}
where
\bea
R^* &\simeq &\left(-m^*_DM^{-1}, \quad  m^*_D(M_R^\dagger)^{-1}\right), \quad  M\equiv M^*_R\mu_X^{-1}M_R^{\dagger}, \label{hue1}\\
m^*_DM^{-1} m^\dagger_D&\simeq & m_{\nu}\equiv U^*_{\mathrm{PMNS}}\hat{m}_{\nu}U^{\dagger}_{\mathrm{PMNS}},  \label{hue2}\\
V^* \hat{M}_N V^{\dagger}& \simeq & M_N+ \fr{1}{2}R^TR^* M_N+ \fr{1}{2} M_NR^{\dagger} R,\; M_N\equiv \begin{pmatrix}
	0&M_R^*  \\
	M_R^{\dagger}& \mu_X
\end{pmatrix}
.  \label{hue3}
 \eea 
The relations between the flavor  and mass eigenstates are
\be
\nu'_L=U^{\nu} n_L, \quad  \mathrm{and} \; (\nu'_L)^c=U^{\nu*}  (n_L)^c, \label{Nutrans}
\ee
where $n_L^T\equiv(n_{1L},n_{2L},...,n_{9L})$ and $(n_L)^c\equiv((n_{1L})^c,(n_{2L})^c,...,(n_{9L})^c)^T$.
 The standard form of the lepton mixing matrix $U_{\mathrm{PMNS}}$ is a function   of three angles $\theta_{ij}$, one Dirac phase $\delta$ and two Majorana phases $\al_{1}$, and $\al_2$, ~\cite{pdg}, 
 namely
\begin{align}
	U_{\mathrm{PMNS}}
	&= \begin{pmatrix}
		1	& 0 &0  \\
		0	&c_{23}  &s_{23}  \\
		0&  	-s_{23}& c_{23}
	\end{pmatrix}\,\begin{pmatrix}
		c_{13}	& 0 &s_{13}e^{-i\delta}  \\
		0	&1  &0  \\
		-s_{13}e^{i\delta}&  0& c_{13}
	\end{pmatrix}\,\begin{pmatrix}
		c_{12}	& s_{12} &0  \\
		-s_{12}	&c_{12}  &0  \\
		0& 0 	&1
	\end{pmatrix} \mathrm{diag}\left(1, e^{i\alpha_{1}},\,e^{i\alpha_{2}}\right) \label{pt34m}
	\crn
	&=U^0_{\mathrm{PMNS}} \;\mathrm{diag}\left(1, e^{i\alpha_{1}},\,e^{i\alpha_{2}}\right),
\end{align}
where $s_{ij}\equiv\sin\theta_{ij}$, $c_{ij}\equiv\cos\theta_{ij}$,  
 $i,j=1,2,3$ ($i<j$), $0\le \theta_{ij}<90\; [\mathrm{Deg.}]$ and $0<\delta\le 720\;[\mathrm{Deg.}]$. The Majorana phases are chosen in the range $-180\le\alpha_i\le 180$ [Deg.]

The 331ISS model inherits interesting 
 consequences in the neutrinos sector that does not exist 
 in other 3-3-1 models with $\beta \neq \fr{1}{\sqrt{3}}$.  Particularly, the Dirac mass matrix $m_D$ is always antisymmetric, hence predicts one massless active neutrino at tree level.  In addition, $m_D$ cannot be 
 parameterized using the most popular form introduced in Ref. ~\cite{Casas:2001sr}.  Instead, $m_D$ has a very special form ~~\cite{in331}, leading to  strict relations of   cLFV rates  if all heavy neutrinos have the same masses, such as  Br$(\mu \to e\gamma)/$Br$(\tau \to \mu \ga)=$ constant and depends precisely on the neutrino oscillation data. On the other hand, non degenerate heavy neutrino masses can relax this cLFV relations, but the recent experimental constraint of Br$(\tau \to \mu \ga)$ results in the largest  value of the deviation of $(g-2)_{\mu}$ anomaly from the SM is around $10^{-9}$, not enough to explain the experimental $(g-2)_{\mu}$ data ~\cite{in332}. A solution for this problem is adding a new singly charged Higgs boson so that new Yukawa couplings will lead to new one-loop contributions that cancel the large cLFV amplitudes, while increase the $(g-2)_{\mu}$ value. The allowed parameter regions explaining successfully the  $(g-2)_{\mu}$ experimental data were shown in Ref.  ~\cite{in332}.  A study  discussing on the regions explaining both experimental data of $(g-2)_{e,\mu}$ anomalies as well as LFV decays of charged leptons and the SM-like Higgs boson was introduced ~~~\cite{Hong28}. Finally, the 331ISS model has rich physical consequences in the lepton sector because of the very special form of $m_D$, which is deserved for further studies other LFV decays such as $Z\to l_b l_a$, $l_b\to l_a l_c l_d$, etc. 

\section{The  3-3-1 model with non-Abelian discrete symmetries}
\lb{sec4}
It is well known that neutrinos are massive  
and their mixing with very special forms. A prospective research direction has been studied extensively, to explain lepton mixing pattern, the small quark mixing angles, the tiny masses of neutrinos and other important phenomenologies, that is the combination of discrete symmetries and the SM or its extensions. In this treatment, discrete symmetries, such as $A_4, S_3, S_4, D_4, T_7$ and so on, have been added to the BSMs 
such as the 3-3-1 models, $B-L$ models, etc 
 ~\cite{A41,A42,S31,S41,D41,T71}. On the other hand, the most natural solution of the quark and lepton mass hierarchies probably  are  by the Froggatt-Nielsen mechanism or by non-Abelian discrete symmetries ~\cite{Froggatt79,Leurer93,Vien22}. Within the Froggatt-Nielsen mechanism, the obtained values of the three neutrino mixing angles are quite small compared to the experimental data, and while the quark and charged-lepton mass hierarchies can be explained within this mechanism, the specific predictions suffer from relatively large errors.
 	
 Apart from neutrino phenomenology, the 3-3-1  models also naturally accommodate potential candidate for DM which have been applied with discrete symmetry ~\cite{Longprd19,Montero18,Loi21}.
Moreover, in recent years, the anomalous magnetic moment (AMM) of charged leptons being an interesting problem that goes BSMs. 
There is an inconsistency in AMM
 between theoretical and experimental values. The experimental magnitude of the differences owns both negative and positive signs, with $\De a_e= (-8.8 \pm 3.6)10^{-13}$ ~~\cite{Parker18,Isha23} and $\De a_e=(4.8 \pm 3.0)10^{-13}$ ~~\cite{Morel20,Isha23}. Therefore, we need further investigations of electron's  AMM  (ae)
  to confirm the sign of $\Delta a_e$. Further, 
  the   muon's AMM  ($a_\mu$) plays a central role in precision tests of the SM. The difference between experiment and theory on $a_\mu$ is determined as $\De a_\mu = (251 \pm 59)10^{-11}$ ~\cite{Isha23}. The difference between theoretical and experimental values of the electrons's and muon's AMM
  will open the great prospects for physics beyond the SM which can be addressed within the framework of the 3-3-1 model ~~~\cite{Antonio33121} and the 3-3-1 model with discrete symmetry ~\cite{Antonio331d4}.

\section{The  3-3-1 model with cosmological inflation}
\lb{sec5}
The axion is very attractive to experimental searches, is a popular topic in the modern physics. This is arised from the spontaneous breaking of the the global $U(1)_{Q_A}$ symmetry that was implemented by Peccei - Quinn (PQ) to solve the Strong $CP$  
problem ~\cite{pq1,pq2}.. 
 
\subsection{\label{pqc} The 3-3-1 model with axion like particle}
\lb{sec51}
In the frameworks of the 3-3-1 models, the PQ formalism was  considered for
in the version with  $\bet = -\fr 1{\sqrt{3}}$, almost two decades ago
 ~\cite{a1,a2,a3}. The main ingredient is a singlet scalar $\phi$ carrying lepton number two.
However, in the above  mentioned works, the singlet $\phi$ is expanded  as
\be \phi =  \fr{ 1}{ \sqrt{2}} (v_\phi + R_\phi + i I_\phi ) \,.
\lb{may261} 
\ee 
It is worth mentioning that within above expansion of scalar $\phi$, its pseudoscalar $I_\phi$ should be called
ALP but not axion. 
In this case, the ALP mixes with other CP odd scalars. The axion is the pure imaginary
part of $\phi$ only in the limit $v_\phi \gg v_\chi$, which is  the VEV responsible for breaking from $SU(3)_L$  to SM subgroup. 
In Ref. ~\cite{a3}, the PQ symmetry was considered for two main versions:  M331  and  331RN  models~\cite{julio1,julio2}.

New development in this direction was done seven  years ago in Ref. ~\cite{jpf}, where the discrete symmetry $Z_{11}\times Z_2$ is imposed, and Majorana right-handed neutrinos are introduced. To provide  mass of Majorana neutrinos, a complex scalar transforming as a $ \mbox{SU(3)}_L$  singlet $\phi$ is added. As a consequence, the model also contains a heavy $CP$ even scalar with mass
in the range of $10^{10}\, \gev$ identified  as an inflaton. However,  Ref. ~\cite{jpf} still contains some flawed points  
such as no identification of 
the SM-like Higgs boson and incorrect mixing of $CP$ odd scalars.  The  mentioned problems
have been recently solved in Ref. ~\cite{alp331}. 
 As a result, the model contains the expected inflaton with mass around $10^{11} \, \gev$, a heavy scalar with  mass at TeV scale labeled by $H_\chi$, one scalar with mass at the EW scale labeled by ($h_5$), and of course the SM-like Higgs boson ($h$).
	
The particles and their transforms ~\cite{jpf,alp331} under  $\mbox{SU(3)}_C\times \mbox{SU(3)}_L\times \mbox{U(1)}_N \times Z_{11}  \times Z_2$ group
are presented  in  Table \ref{tabm1}.
	\small{
		\begin{table}[th]
			\resizebox{8cm}{!}{
				\begin{tabular}{|c|c|c|c|c|c|c|c|c|c|c|c|c|c|c|}
					\hline
			& $Q_{3 L}$ & $Q_{nL}$ & $u_{a R}$ & $d_{a  R}$ & $T_{3R}$ &$ D_{n R} $ & $\psi_{aL}$ & $
					l_{aR}$ & $N_{aR} $ & $\eta$ & $\chi $ & $\rho$ & $
					\phi$  \\ \hline
					$SU(3)_C$ & $\mathbf{3}$ & $\mathbf{3}$ & $\mathbf{3}$ & $\mathbf{3}$ & $%
					\mathbf{3}$ & $\mathbf{3}$ & $\mathbf{1}$ & $\mathbf{1}$ & $\mathbf{1}$ & $%
					\mathbf{1}$ & $\mathbf{1}$ & $\mathbf{1}$ & $\mathbf{1}$   \\ \hline
					$SU(3)_L$ & $\overline{\mathbf{3}}$ & $\mathbf{3}$ & $\mathbf{1}$ & $%
					\mathbf{1}$ & $\mathbf{1}$ & $\mathbf{1}$ & $\mathbf{3}$ & $\mathbf{1}$ & $%
					\mathbf{1}$ & $\mathbf{3}$ & $\mathbf{3}$ & $\mathbf{3}$ & $\mathbf{1}$  \\ \hline
					$U(1)_N$ & $\fr 1 3 $ & $0 $ & $\fr 2 3$ & $-\fr 1 3$ & $\fr{2
					}{3}$ & $-\fr 1 3$ & $-\fr 1 3$ & $-1$ & $0$ & $-
					\fr 1 3$ & $-\fr 1 3$ & $\fr 2 3$ & $0$
					\\ \hline
$Z_{11}$ & $\om_0$ & $  \om^{-1}_4$ & $\om_5$ & $\om_2$
					& $\om_3$ & $\om_4$ & $\om_1$ & $\om_3$ & $%
					\om^{-1}_5$ & $\om^{-1}_5$ & $\om^{-1}_3$ & $\om^{-1}_2$ & $\om^{-1}_1$  \\
					\hline
$Z_2$ & $1$ & $1$ & $-1$ & $-1$ & $1$ & $1$ & $1$ & $-1$ &
					$-1 $ & $-1$ & $1$
					& $-1$ & $1$ \\ \hline
					\hline
			\end{tabular}}\vs
			
			\caption{$SU(3)_C\otimes SU(3)_L\otimes U(1)_N\otimes Z_{11}\otimes Z_2$ charge assignments of the particles in  the model.
				Here $w_k=e^{ik2\pi/11}$, $a=1,2,3$ and $\al =1,2$.}
			\label{tabm1}
	\end{table}}

Masses of  fermions and gauge bosons request  VEVs of three triplets and one singlet
\bea \sqrt{2}\, \langle \eta \rangle^T  & = & \left( v_\eta , 0, 0\right)\,, \,
\sqrt{2}\, \langle \chi \rangle^T   =  \left( 0 , 0, v_{\chi}\right) \,,
	\crn
	\sqrt{2}\, \langle \rho \rangle^T & = &  \left( 0, v_\rho, 0
	\right) , \,\hs
	\sqrt{2}\, \langle \phi \rangle  =  v_\phi\,.
	\label{eq2t}\eea
The symmetry breaking of the model under consideration    is in three steps by following scheme:
\bc
	\begin{tabular}{c}  
$SU(3)_C\otimes SU(3)_L\otimes  U(1)_X \otimes Z_{11} \otimes Z_2  \otimes U(1)_{Q_A}$\\
$  \downarrow \langle \phi \rangle $\\
$SU(3)_C\otimes SU(3)_L \otimes U(1)_X  \otimes Z_2$\\
$\downarrow \langle \chi  \rangle $\\ 
$SU(3)_C\otimes  SU(2)_L \otimes U(1)_Y$\\
$\downarrow v_\eta, v_\rho$\\
$SU(3)_C\otimes  U(1)_Q.$
\end{tabular}
\ec 
 The version with  axion ~\cite{lh331}  is presented in the next section.
     
\subsection{\label{pqc2} The 3-3-1 model with axion}
\lb{sec52}
The new format of writing PQ transforms is given in Refs. ~\cite{giogi,jekim,luzio,gu,choi}.
For future presentation, here  we  write explicitly  the scalar $\phi$ in the form 
	\be 
	\phi =  \fr 1 2 (v_\phi + R_\phi)\,  e^{i \fr{a}{2 f_a}c_\phi} \,.
	\label{eq2}\ee
According to  Ref. ~\cite{a2}, and using notations in Refs. ~\cite{gu,choi} 
for an arbitrary fermion and scalar boson,  the PQ  transformations are as  follows
\bea
f & \rightarrow & f^\prime =  e^{ i  \left(\fr{c_f}{2 f_a}\right)\ga_5 a} f\,, \hs {\bar f} \rightarrow {\bar f}^\prime =  {\bar f} e^{ i  \left(\fr{c_f}{2 f_a}\right)\ga_5 a} \,, \hs
\va \rightarrow \va^\prime = e^{ i \left(\fr{c_\va}{2 f_a}\right) a} \va \,.
\label{pqr}\
\eea
For  chiral fermions in the model under consideration, this is
\bea
 f_L &\rightarrow& f^\prime_L  =  e^{ - i  \left(\fr{c_f}{2 f_a}\right) a} f_L\,,\hs 
{\bar f}_L \rightarrow {\bar f}^\prime_L  = {\bar f}_L  e^{  i  \left(\fr{c_f}{2 f_a}\right) a}\,,\crn
 f_R &\rightarrow & f^\prime_R  =  e^{  i  \left(\fr{c_f}{2 f_a}\right) a} f_R\,,\hs 
{\bar f}_R \rightarrow {\bar f}^\prime_R  = {\bar f}_R  e^{ - i  \left(\fr{c_f}{2 f_a}\right) a}\,,  \label{hay1}
\eea
where $c_f$ is  PQ charge of fermion and $f_a \sim 10^{11} \gev$  is  axion decay constant related  to
the scale of symmetry breaking of $U(1)_{PQ}$ global group. 
The values
 $c_F$ for transformations of fermions under the $Z_{11}$ symmetry are given as ~\cite{jpf}
\be  
  c_u = c_T =  - c_d = - c_D =   c_l = - c_{l R}  = - c_\nu =  c_{\nu_R} = - c_N  \equiv R \, .  
\label{s1}
\ee
 For scalars, from Yukawa couplings, it follows that  charged scalars have vanishing   PQ charge  
 since they connect {\it up and down}  particles with 
opposite values,    while electrically  neutral scalar has  PQ charge duplicate charge  (with opposite sign)  of fermion to which it provides mass,  because  it  connects to {\it both up or down} particles: 
\bea  
\eta_1^0 & \rightarrow & e^{ i \fr{a}{ f_a}} \eta_1^0\,, \hs \chi_1^0
\rightarrow e^{ i \fr{a}{ f_a}} \chi_1^0 \, ,\hs \rho_1^+
\rightarrow \rho_1^+\,,\crn 
\phi  & \rightarrow &   e^{i \fr{a}{ f_a}} \phi \,, \hs    \rho_2^0
\rightarrow e^{- i \fr{a}{ f_a}} \rho_2^0 \, ,\hs  \chi_2^-
\rightarrow \chi_2^-\,.
\label{s3}
 \eea  
The  term $ \la_\phi\ep^{ijk} \eta_i \rho_j \chi_k \phi^*$
 shows that $\rho$  has  the same value and opposite in sign to that of $\eta$,  $\chi$ and $\phi$ ~\cite{lh331}. This
is  explicitly  clarified in Table \ref{tab2}.  Hereafter, the PQ charge is renamed by $Q_A$ charge.
\begin{table}[th]
\resizebox{12cm}{!}{
				\begin{tabular}{|c|c|c|c|c|c|c|c|c|c|c|c|c|c|c|c|c|c|c|}
					\hline
					&  $\, u \, $ & $\, d\, $ & $T$ & $D_\al$ & $\, l\, $ &$ \nu$ & $\nu_R$ & $
					N_R$ & $\eta_1^0 $ & $\eta_3^0$ & $\chi_1^0 $ & $\chi_3^0$ & $
					\rho^0$ &$\phi$&$\eta_2^-$&$\chi_2^-$&$\rho_1^+$&$\rho_3^+$ \\ \hline
					$U(1)_{Q_A}$ & $1$ & $-1$ & $1$ & $-1$ & $
					-1$ & $+1$ & $1$ & $-1$ & $2$ & $%
					2$ & $2$ & $2$ & $-2$  & $2$  & $0$  & $0$  & $0$  & $0$ 
					\\ \hline
			\end{tabular}}\vs 
			
			\caption{$U(1)_{Q_A}$ charge assignments of the particle content of the model.
				Here $Q_A(F)=c_F = c_{F_R} = - c_{F_L}$}
			\label{tab2}
	\end{table}
	
Note that the Higgs sector in  this model is  similar to that in   Ref. ~\cite{alp331} with just difference
in the $CP$ odd sector, where the component $I_\phi$ is decoupled.

In the limit $v_\phi \gg v_\chi \gg v_\rho \gg v_\eta $, the scalar content of the model can be presented as follows ~\cite{alp331}
		\bea
\rho & \simeq &
		\left(
		\begin{array}{c}
			G_{W^+} \\
			\fr{1}{\sqrt{2}}\left(  v + h+ i G_Z\right)\\
			H_2^+ \\
		\end{array}
		\right)\,,\,	
		\eta  \simeq  \left(
		\begin{array}{c}
			\fr{1}{\sqrt{2}}\left(  u +h_5 + i A_5\right) \\
			H_1^- \\
			G_{X^0}   \\
		\end{array}
		\right),\, \, \,	
			\chi  \simeq 
		\left(
		\begin{array}{c}
			\chi_1^0 \\
			G_{Y^-} \\
			\fr 1{\sqrt{2}}\left( v_\chi  + H_\chi + i G_{Z'}\right) \\
		\end{array}
		\right)\, ,\,\,
		\crn
		\phi & = & \fr{1}{\sqrt{2}}\left( v_\phi + \Phi\right) e^{- i \fr{ a}{\, f_a}}\, .
		\label{d25}\eea
Combination of Table \ref{tabm1} and   
Eq. \eq{d25} leads to the following interesting consequences: 
 Firstly, the SM-like Higgs boson $h$  has Yukawa couplings with only  SM fermions. Secondly, the heavy scalar  $H_\chi$  and pseudoscalar  $A_5$  can have Yukawa couplings with not only exotic quarks but also SM quarks and leptons.
 
For singlets of right-handed fermions and scalar $\phi$, the generators $T_3$ and $T_8$ produce  the zero, but  the general PQ charge $\mathcal{X}_{pq}$ for right handed fermions  takes opposite value 
of left-handed partner, while
for $\phi$, as usually,  it is followed   from Yukawa couplings, i.e., $\mathcal{X}_{pq}(f_R) = - Q_A(f_L)$
 and   $\mathcal{X}_{pq}(\phi) =  2\,, \Rightarrow Q_A(\phi) = 2$. 
 To be noted that there exists the same situation when one deals with electric charges  of right-handed fermions.
 
Note that the parameter $R$ in Eq. \eqref{s1} in principle  is any non-zero integer. However, 
as seen later, within notations in Eq. \eqref{pqr},  the absolute value of $R$ should be equal to the unity, i.e.,
 $\vert R \vert = 1$. From kinetic part of scalar $\phi$, it follows
 \bea  
 \left(\fr{c_\phi}{2 f_a }\right)^2  & = &\left(\fr 1{f_a}\right)^2 \,, \label{lp1}\\
 f_a & = & v_\phi\,.  \label{lp2}
 \eea
  From \eq{lp1}, one gets $c_\phi= \pm 2$. From \eq{s1} and \eq{s3}, one obtains $c_\phi = 2 c_u$. Therefore  $\vert R \vert = 1$, and
 we choose the sign is plus, i.e., $R = 1$. 
 
{\bf i) Formula for PQ charge} 

The PQ charges given in Table \ref{tab2} allows us to write some nice formula similar to  generalized lepton number in ~\cite{longchang,jhep18}. 
From Table \ref {tab2}, 
and Eq. \eqref{s3}, we can formulate  PQ charge  assignment for triplets in the model under consideration
as follows 
\be 
Q_A =  2 \,  T_3 - \fr 2{\sqrt{3}}\,  T_8  +  \mathcal{C}_{f}\,.
\label{s161}
\ee
Let us call $\mathcal{C}_{f}$ by hyper-chirality. It is worth mentioning that the formula \eq{s161} is firstly constructed for BSMs.
 Looking at  Eq. (9) in Ref. ~\cite{julio2}, one can see that our formula is much better.

 Therefore, the PQ transformation in Eqs. \eqref{hay1} and \eqref{s3} can be written in form of PQ charge operator labeled 
 by $Q_A$ as follows 
\bea 
U(1)_{Q_A}:  \hs 
f & \rightarrow & f^\prime =  e^{ i  \left(\fr{a}{2 f_a}\right)\ga_5 Q_A} f\,, 
\hs {\bar f} \rightarrow {\bar f}^\prime =  {\bar f} e^{ i  \left(\fr{a}{2 f_a}\right)\ga_5 Q_A} \,, \label{s163t}\\
 f_L &\rightarrow& f^\prime_L  =  e^{ - i  \left(\fr{a}{2 f_a}\right) Q_A} f_L\,,\hs
{\bar f}_L  \rightarrow  {\bar f}^\prime_L  = {\bar f}_L  e^{  i  \left(\fr{a}{2 f_a}\right) Q_A}\,,\crn
 f_R &\rightarrow & f^\prime_R  =  e^{  i  \left(\fr{a}{2 f_a}\right) Q_A} f_R\,,\hs 
{\bar f}_R \rightarrow {\bar f}^\prime_R  = {\bar f}_R  e^{ - i  \left(\fr{a}{2 f_a}\right) Q_A}, \, 
\label{s162}\\
\va &  \rightarrow & \va^\prime  =  e^{ +i  \left(\fr{a}{2 f_a}\right) Q_A} \va \,. \label{s163}
\eea
 It is worth noting that in the realization  the Georgi-Kaplan-Randall (GKR) field basis,
  all fields  except the axion, transform by additive  constant. It can be proved that the two transformations 
  of $f_L$ and $f_R$ result from the combined transformation of $f=(f_L,f_R)^T$ given in Eq. \eqref{s163t}. Consequently,
   a four-component fermion $f$ and its right-handed component have the same $Q_A$ value and always have a
    opposite sign with the left-handed component. 
    
It is important to note that the PQ charge of the similar singlet $\si$ in Ref. ~\cite{julio2} cannot be  fixed
 (see Table 1 there).

 The following remarks are in order.
\ben
\item  The formula \eqref{s161} shows that  the difference of electric charges of up and down quarks/leptons  is 1, i.e., $\De Q =1$,  while for PQ charge $\De Q_A =2$. 
 For fermions, electric charges ($Q$) of left-handed and right-handed fermions are to be equal, while their PQ ($Q_A$) charges are opposite.  For scalars,
  only neutral scalars have   PQ  charges equal to $\pm 2$, where the sign plus and minus correspond to  scalars in top and bottom of doublets. The charged scalar does not have a PQ charge. So the electric charge ($ Q $)  and ($ Q_A $) have  again opposite property.
\item It is worth mentioning that in the model under consideration, the relations among PQ charges ($Q_A$) are quite simple
\be
Q_A(u) = - Q_A(v) = \fr 1 2 \left[ Q_A(\eta) = Q_A(\chi) = - Q_A(\rho) = Q_A(\phi)  \right] = 1\,,
\label{dec30}
\ee
which is completely different from those in ~\cite{julio2}.
\een  
  
To solve the strong $CP$ problem, the  $U(1)_{Q_A}$ symmetry is spontaneously breaking by  VEV  of the singlet $\phi$ as follows
\be Q_A \langle \phi \rangle =  \sqrt{2}\,  v_\phi \neq 0\,.
\label{s153}
\ee
From Table \ref{tab2}, we see that PQ charge of neutral scalar equals twice of PQ charges of fermions receiving for which scalar provides masses, so all  Yukawa    couplings 
 are invariant.   
 In addition, exotic quarks carry lepton number 2, while ordinary SM quarks 
 do not, so mass eigenstates of exotic quarks are their original states, while  ordinary quarks are with mass mixing. 
 
{\bf  ii) Axion couplings} 

- The axion - fermion  derivative  couplings  have the form
\bea
\mathcal{L}_ {(f-a)} & = &  + \left(\fr{1}{ \, f_a}\right) \pa^\mu \, a \,\left[ \bar{d}\,  {\bf c}_{d} \,\ga_\mu 
 \ga_5 d + \bar{u}\, {\bf c}_{u}\,\ga_\mu  \ga_5 u   + \bar{T}\, {\bf c}_{T}\,  \ga_\mu  
 \ga_5 T + \bar{D}_\al \, {\bf c}_{D_\al} \, \ga_\mu   \ga_5 D_\al \right. \label{s8}\\
 && + \left. \bar{l}\,  {\bf c}_{l}\ga_\mu \, \ga_5 l  + I_{\nu}\bar{\nu}_a \, 
 {\bf c}_{\nu}\, \ga_\mu  \ga_5 \nu_a + \fr 1 2  \bar{N}_{a}\, {\bf c}_{N_a} \, \ga_\mu  P_R  N_{a}  
 \right]\,.
 \label{s9}
\eea
In  Eq. \eqref{s9}, for the  coefficients  (${\bf c}_{f}, f= d,u,\cdots N_R$), one has to  count  the number of color, flavor indexes and PQ charge $Q_A(f)$.  

-  Regarding the scalar sector,   Lagrangian  for covariant kinetic terms of a complex scalar $\va$  is
\bea \mathcal{L}_\va & = & (D^\nu \va)^\dag D_\nu \va  = [(\pa_\nu - i P^\va_\nu )\va]^\dag (\pa^\nu - i
 P^{\va \nu })\va\crn
& = & \pa^\nu \va^\dag \pa_\nu \va  - i \pa_\nu \va^\dag 
P^{\va \nu } \va + i \va^\dag  P^{\va \nu} \pa_\nu \va 
+ \va^\dag   P^\va_\nu   P^{\va \nu} \va.
 \label{s10}
\eea
The total part concerned to axion is given below
\bea
\mathcal{L}_a & = & \fr 1 2 \pa_\nu a \pa^\nu a - \fr 1 2 m_{a o }^2 \, a^2 
 +  c_{G G} \fr{\al_s}{4 \pi} \fr{a}{2 f_a} G \tilde{G} + 
c_{WW}\fr{\al_2}{4 \pi} \fr{a}{2 f_a} W \tilde{W}
+ c_{BB}\fr{\al_1}{4 \pi} \fr{a}{2 f_a} B \tilde{B}
\label{s181}\\
& + &  \fr{\pa^\al a}{2 f_a} \left(\sum_{f=u,d,T,D}^{l,\nu}  \bar{\psi}_f c_f \ga_\al \ga_5 \psi 
 + \fr 1 2  \bar{N}_{a}\, {\bf c}_{N_a} \, \ga_\al  P_R  N_{a} \right) - 
  \left(\bar{q}_L\,  M_q\,  q_R + \mbox{H.c.}\right)\label{s182}\\
& + &  \left(\fr{1}{ f_a}\right)^2  \pa_\al a \pa^\mu a \left(\sum_{H= \eta_1^0, \eta_3^0,
\rho_2^0}^{ \chi_1^0, \chi_3^0, \phi} H^* H\right)\hs \hs ;  \label{s183}\\
&-& i \left(\fr{c_\va}{ 2 f_a}\right) \pa^\al a \sum_{D=\eta,\chi, \phi}^{K=\rho^0} 
\left[D^*\stackrel{\leftrightarrow}{\pa_\al} D - K^*\stackrel{\leftrightarrow}{\pa_\al} K\right]
\label{s184}\\
& + & 2 \left(\fr{c_\va}{2 f_a}\right)  \pa^\al a\,\sum_{H= \eta_1^0, \eta_3^0,
\rho_2^0}^{ \chi_1^0, \chi_3^0, \phi}  H^\dag P_\al^{H } H
\,,\label{s185}
\eea   
where $H^* H= \fr{1}{2}[(v_H + R_H)^2 + I_H^2]$.  It is worth emphasizing   that in \eq{s182} the matrices   $ M_q  $ are diagonal. 

{\bf iii) Axion mass}

In the chiral perturbative theory, the axion mass arises from the part ~\cite{chp2,chp3}
\bea 
\mathcal{L}_{amass } & = &  \fr{f_\pi^2}{4} 2 B_0 2 \fr 1 3 (m_u+m_d+m_s) \cos  \fr{\mathcal{K}}{f_\pi} \crn
& =& \fr 1 6 f_\pi^2 
( m_\pi^2 + m_{K^0}^2 +  m_{K^+}^2 )\cos  \fr{\mathcal{K}}{f_\pi} -\fr{m_a^2}2\,.
\label{oct42}
\eea
Here  ~\cite{lh331}
\bea
m_a^2  &  = &   \fr{1}{2}
B_0   \left(\fr{f_\pi}{ f_a}\right)^2 \left(\fr{m_u m_d m_s}{m_u  m_d + m_u  m_s +m_d  m_s }\right)\crn
&  = &   \fr{1}{2} m_\pi^2
\left(\fr{f_\pi}{ f_a}\right)^2 \fr{1}{(m_u + m_d)}\left(\fr{m_u m_d m_s}{m_u  m_d + m_u  m_s +m_d  m_s }\right) 
\nn
\eea
This result  is similar to that in ~\cite{sr}
\bea m_a &=& 4  \fr{f_\pi m_\pi}{f_a/N}\left[\fr{ m_u m_d m_s}{(m_u m_d + m_u m_s + m_d m_s)(m_u + m_d)}  \right]^{\fr 1 2}\crn
&\simeq & (1.2 \times 10^{-5} \, \textrm{eV})\left( \fr{10^{12} \, \gev}{f_a/N} \right)\,,
\label{s121}
\eea

\section{Phenomenology}
\lb{s6}
 In this section, I just mention on main  consequences of the models.
\subsection{Collider physics}
\lb{sec61}
The doubly charged  beleptons $X^{\pm \pm}$ of the minimal version are highly attached and their production
at LHC was considered in ~~\cite{lhc1t}. Production of the Higgs boson in the M331 model
at the $e^+ e^-$ Next Linear Collider, and  in CERN Linear Collider (CLIC) has been considered 
in ~\cite{7}.
The   bilepton gauge boson  masses are usually fixed to be in the range of 600 GeV, while mass of the extra neutral gauge boson $Z^\prime$ is limited
in the region  [1 $\div $ 1.2] TeV  ~\cite{lhc1}.

The limits on $Z^\prime$ mass in the 331RN model 
is as follows:  
$ M_{Z^\prime} > 4.1$ TeV ~\cite{lhc2t}.

At the LHC, the cross section of $pp\to Z^\prime \to XY$ is 
\be
\si_{(pp\to V'\to\bar{f}f')}=\sum_{\{ij\}}\int_{\tau_{0}}^{1}
\fr{d\,\tau}{\tau}\cdot\fr 1 s \fr{d\,\mathcal{L}_{ij}}{d\,\tau}\cdot[\hat{s}\,\hat{\si}_{(ij\to V'\to\bar{f}f')}(\hat{s})]\,,
\ee 
where $X$ and $Y$ denote the decay products of the $Z^\prime$ boson,
$\sqrt{s}$ is the total energy of the incoming proton-proton beam,
$\sqrt{\hat{s}}$ is the partonic center-of-mass  energy and $\tau\equiv \hat{s}/s$
and $\fr{1}{s}\fr{d\,\mathcal{L}_{ij}}{d\,\tau}$ is the parton lumonosity ~\cite{lhc0}.

It is worth noting that the single $Z^\prime$ can be produced in the hadron  collision.
 The total
cross-section of the scattering process $p p\to Z^\prime$ is
given by ~\cite{lhc1}
 \bea \si( p p \to Z^\prime
)&=&2\sum_{i=1}^5\int_0^1dx_1\int_0^1dx_2f(i,x_1,Q)f(-i,x_2,Q)\hat\si(q_i\bar
q_i\to Z^\prime )\crn &=&\fr{2\pi
g^2}{3c^2_W}\fr 1 s \sum_{i=1}^5[(g_{2V}^{qi})^2+(g_{2A}^{qi})^2]
\int_{M_{Z^\prime}^2/s}^1f(i,x_1,M_{Z^\prime})f\left(-i,
\fr{M_{Z^\prime}^2}{sx_1},M_{Z^\prime}\right)\fr{dx_1}{x_1}.\label{totalcr}
\eea 
where 
 \bea \hat\sigma(q\bar q\to Z^\prime
)=\fr{1}{3}\fr{\pi g^2}{c^2_W}[(g^q_{2V})^2+(g^q_{2A})^2]\delta
(\hat s-M_{Z^\prime}^2)\, ,\hs   s =x_1x_2s. \eea
This subprocess in the M331 
 model  was considered in ~\cite{14}.

For hadron  collider CERN  LHC 
  the suitable references
 are  ~\cite{lhc1,lhc2,lhc3}, for FCC  ~\cite{eec1,eec2,eec3}. 

\subsection{Neutrino mass and mixing}
\lb{sec62}

In this subsection, the mechanisms  for 
reproducing the light neutrino mass in the frameworks  of  two pioneer  3-3-1 gauge models (M331 model and 331RH)
  are briefly reviewed. It is worth noting that, at the tree level, neutrinos are massless in both versions. 

a) {\it Generation of neutrino mass in  the minimal 3-3-1 model} 

Due to the  Landau pole around 5$\tev$,
the  five-dimensional  operator:
$ h(\bar f^C_L \eta^*)(\eta^{\dagger}f_L)/\La $ provides  mass 
 to the neutrinos $m_\nu= h v^2_\eta/\La$.
Then, for $v_\eta \approx 10^2 \gev$ and $\La=5$TeV, we get $m_\nu= 10\, h \gev$
As a consequence,  the effective dimension-5 operator cannot provide 
an adaptable neutrino mass. 

Regarding the tiny   neutrino mass issue  in the  M331 model, to 
overpass  this trouble  
 there are two ways. The first one is combination of 
   symmetries and effective dimension-11 operator. As a result, neutrinos get mass 
5 to get neutrino 
at $eV$  scale ~\cite{neumass1t}.
  The second outlet is adding  
of singlet  RH    neutrinos 
    to the leptonic particle content of the  M331 model 
    and then  combination of  type I and type II seesaw mechanisms leads to 
   small  neutrino masses~\cite{neumass1}.

b) {\it Generation of neutrino mass in  the  3-3-1 model with RH neutrinos} 

In the original version of the model,  neutrinos are massless. Addition of 
the following  effective dimension-five operator,
\be 
{\mathcal L}_{M_L}=	\fr{f^{ab}}{\Lambda}\left( \overline{L^C_a} \eta^* \right)\left( \eta^{\dagger}L_{b} \right)+\mbox{H.c}.
	\label{5dL}
\ee 
leads to the left-handed neutrinos develop  Majorana mass terms $
(M_{L})_{ab}= f_{ab}v^2_\eta /\Lambda$. 

The Majorana masses for the right-handed neutrinos are followed from coupling
\bea 
{\mathcal L}_{M_R}&&=	\fr{h^{ab}}{\La}\left( \overline{L_{a}^C} \chi^* \right)\left( \chi^{\dagger}L_{b} \right)+\mbox{H.c}.
	\label{5dR}
\eea 
leading to term
$ (M_{R})_{ab}\overline{(\nu_{aR})^C} \nu_{bR}$ with  $(M_{R})_{ab}=h_{ab}v^2_{\chi^{\prime}}/\La $.

The type II seesaw mechanism can be realized by adding scalar sextet $S \sim (1,6,-\fr 2 3)$ leading
to effective coupling  $\bar L SL^C $  ~~\cite{neumassrhn4}.

It is well known that the seesaw mechanism is the best way for  tiny neutrino mass. For this purpose,   type II
seesaw mechanism has been imposed  in the minimal version  ~\cite{neumass1}.  For the version with  $\bet = -\fr 1{\sqrt{3}}$, the tiny neutrino mass is realized by inverse  seesaw mechanism
 ~\cite{neumassrhn1} (For more references
see Refs. ~\cite{neumassrhn2,
neumassrhn3,neumassrhn5,neumassrhn6,neumassrhn7,Pires})

c) {\it Inverse Seesaw Mechanism in the 3-3-1 model with RH neutrinos}

At present, the inverse seesaw (ISS) mechanism is attractive since it 
can act  at $\tev$ scale.
In the 331RN  model 
  three singlet neutral fermions have been added : $N_{a_L}\sim(1,0)$ ~~\cite{Dias:2012xp,moreonISS331}.

Then, Yukawa Lagrangian  takes the form
\be
{\cal L}^Y_{\mbox{ISS}}=g^{ab}\ep^{ijk}\bar{L^C_{a_i}}\rho^*_j L_{b_k} + G^{ab}\bar L_a \chi (N_{b_L})^C + \fr{1}{2} \bar N^C_L \mu N_L + H.c.
\label{yuk5}
\ee 
Then, the mass matrix $M_\nu$ has the form
\be
M_\nu=
\begin{pmatrix}
0 & m^T_D & 0 \\
m_D& 0 & M^T\\
0 & M & \mu
\end{pmatrix}.
\label{mass4}
\ee
Here the $3\times3$ matrices are defined as
\be
 M_{ab}=G_{ab}\fr{v_{\chi_3^{0}}}{\sqrt2}\,, \hs 
 m_{Dab}=g_{ab}\fr{v_\rho}{\sqrt2}\,,
\ee 
with  $M_{ab}$ and $m_{D_{ab}}$ are Dirac mass matrices, with the last term  has  anti-symmetric form. The 
 matrix in Eq.~(\ref{mass4}) is typical  for the ISS mechanism. Note that there are two energy  scales corresponding to $v_{\chi_3^0}$ and $v_\rho $. 
  The third  scale of energy, $\mu$, is 
 specific for  the ISS mechanism 
  lying at KeV scale.

The above  mass matrix   leads to three light Majorana neutrinos given by, 
\be 
m_{light} = m^T_D M^{-1}\mu (M^T)^{-1}m_D,
\label{mlf}
\ee 
together with  six 
neutral leptons having  TeV scale masses.
 For $v_\rho \simeq 10^2 \, \gev$,  $ v_{\chi_3^0}\simeq 10^3 \gev$  and $\mu \simeq 10^{-7} \gev $ , one obtains two light neutrinos with masses around eV.

In Refs. ~\cite{kit1,kit2}, generation of  neutrinos  mass at two-loop radiative mechanism for  the minimal version was presented,  while  for version with  $\bet =- \fr 1{\sqrt{3}}$   in Ref. ~\cite{kit3rn}.

\subsection{Lepton flavor violation }
\lb{sec63}
Since leptons and antileptons lie in the lepton triplet, the  lepton flavor violation (LFV)  arises
in the 3-3-1 models. However, the generalized lepton number $ \mathcal{L}$  ~\cite{longchang,jhep18} is conserved.
Based on general formula given in  Ref.  ~\cite{clfv1}, the  LFV of charged leptons in the 3-3-1 models
have been presented in Refs. ~~~\cite{clfv2,clfv3}.

\subsection{$(g-2)_\mu$ puzzle}
\lb{sec64}

In the frameworks of the 3-3-1 models,  the muon $(g-2)$ was studied in ~\cite{amm1}.   Recent $(g-2)$ experimental value has been updated  in Ref. ~\cite{amm2}, and the deviation from the
 SM prediction is given   by 5.1 $\si$ ~\cite{amm2,amm4}.
Moreover, the 331$\bet$
	 model consisting of 6 new inverse seesaw (ISS) neutrinos, named the  331$\bet$ISS model,
and a singly charged Higgs boson can explain successfully both the $(g-2)_{e,\mu}$ data and the
neutrino oscillation data ~\cite{amm5}. For combination of $(g-2)$ with other aspects, the reader is referred to
Refs.  ~\cite{amm6,amm7,amm8}. 

\subsection{Dark matter}
\lb{sec65}
Self-interacting dark matter (SIDM) were considered in the pioneer versions: it has been discussed in the
frameworks of  the M331 model 
  by D. Fregolente and M. D. Tonasse ~\cite{dm1} and for the {331RH model  by Long
and Lan in ~\cite{dm2}. By adding inert scalar singlets, the reader is referred to Dong's works (for example,
see  Ref.  ~\cite{dm3}).

\subsection{Early Universe}
\lb{sec6}

It is well known that the baryon asymmetry of universe (BAU) is one of the most
long-standing puzzles in Particle Physics.  To solve this puzzle, the key ingredients are
three Sakharov  conditions, which are $B$ violation,
$C$ and $CP$ violations, and deviation from thermal equilibrium

The comm acceptable opinion  for the BAU is that in Early Universe, there exists cosmological inflation
in the period from $10^{-36}$ to $10^{-34}$ s after Big Bang.   In the frameworks of the 3-3-1 models,
the inflation was considered in Refs. ~~\cite{inf1,inf2}.
The electroweak phase transition (EWPT) was studied in Refs. ~\cite{cp1,cp2,cp3}.
In addition, some related problems are leptogenesis and sphalerons, for which in  the frameworks of the 3-3-1 models were condisered in Refs. ~\cite{lepg1,lepg2} and ~~\cite{bau1}, respectively.

\section{Conclusions}
\lb{sec7}
In conclusion, the models based on the $\mbox{SU}(3)_C\otimes \mbox{SU}(3)_L \otimes \mbox{U}(1)_X$ gauge group
contain intriguing features.  The models can give explanations on number of fermion generation, PQ mechanism, neutrino mass and mixing as well as dark matter and baryon asymmetry of the Universe. 

It is worth mentioning that in the frameworks of the model with   $\bet = -\fr 1{\sqrt{3}}$, the PQ charge operator is constructed.  The advancements continue 
to unfold, we encourage staying engaged and keeping theoretical interest.

\section*{Acknowledgements}
This research has received funding from  
 National Foundation for Science and Technology Development (NAFOSTED) under grant number 103.01-2023.45. The author thanks P. V. Dong, L. T. Hue and  V. V. Vien for reading draft and corrections.


\begin{thebibliography}{99}


\bib{pdg} S. Navas {\it et al.} (Particle Data Group), Phys. Rev. D {\bf 110}, 030001 (2024).

\bib{criv} A Crivellin, B.  Mellado, Anomalies in Particle Physics
Nature Reviews Physics (2024),
DOI:10.1038/s42254-024-00703-6,  arXiv:2309.03870. 

\bib{ppf1} F. Pisano and V. Pleitez, Phys. Rev. D \textbf{46}, 410 (1992).
\bib{ppf2} P. H. Frampton, Phys. Rev. Lett.  \textbf{69}, 2889 (1992).
\bib{ppf3} R. Foot, O. F. Hernandez, F. Pisano, and V. Pleitez,  Phys. Rev. D \textbf{47}, 4158 (1993).
\bib{ppf4} D. Ng, Phys. Rev. D 49 (1994) 4805 ,  e-Print: hep-ph/9212284. 
\bib{flt1} M. Singer, J. W. F. Valle and J. Schechter, Phys. Rev. D {\bf 22}, 738 (1980).
\bib{flt2}  R. Foot, H. N. Long and Tuan A.
	Tran, Phys. Rev. D {\bf 50}, 34 (R)(1994) [arXiv:hep-ph/9402243]
\bib{flt3} J. C. Montero, F. Pisano and V. Pleitez, Phys. Rev. D {\bf 47},
	2918 (1993)
\bib{flt4} H. N. Long, Phys. Rev. D {\bf 54}, 4691 (1996).
\bib{flt5}  H. N. Long, Phys. Rev. D {\bf 53}, 437 (1996).
\bibitem{flt6} M. Ozer,  Phys.Rev.D \textbf{54} (1996) 1143. 	
\bibitem{pal} P. B. Pal, Phys.\ Rev.\ D {\bf 52} (1995) 1659. [hep-ph/9411406].
\bib{landau1} A. G. Dias, R. Martinez, and V. Pleitez, Eur. Phys. J. C 39, 101 (2005), hep-ph/0407141.
\bib{landau2} A. Doff, C. A. de S. Pires, 
Nucl. Phys. B 992 (2023) 116254. arXiv:2302.08578 
\bib{hueninh1} L. T. Hue, L. D. Ninh, 
 Mod. Phys. Lett. A, Vol. 31, No. 10 (2016) 1650062
 arXiv:1510.00302 

\bib{hueninh2} A. J. Buras, F.  De Fazio,  
 JHEP 08 (2016) 115 , e-Print: 1604.02344. 
\bib{331bet} A. E. Carcamo Hernandez, R. Martinez and F. Ochoa, Phys. Rev. D 73 (2006) 035007 [hepph/0510421].
\bib{e331p} W. A. Ponce, Y. Giraldo and L. A. Sanchez, Phys. Rev. D 67, 075001 (2003).  
\bib{e3311} P. V. Dong, H. N. Long, D. T. Nhung and D. V. Soa,
  Phys. Rev.  D  {\bf 73}, (2006) 035004,  [arXiv: hep-ph/0601046]. 
\bib{dlahep}  P. V. Dong and H. N. Long, 
  Advances in High Energy Physics, {\bf  2008},  739492 (2008), arXiv:0804.3239.
\bib{e331qm1}  P. V. Dong, D. T. Huong, Tr. T. Huong, H. N. Long
 Phys.Rev.D74:053003,2006, arXiv:hep-ph/0607291
\bib{e331qm2} J. C. Montero, B. L. Sánchez-Vega, 
 Phys. Rev. D 84, 055019 (2011), arXiv:1102.5374 .
\bib{e331qm3} P. V. Dong, H. T. Hung, H. N. Long
Phys. Rev. D 86, 033002 (2012), arXiv:1205.5648 .
\bib{e331n} P. V. Dong, D. Q. Phong, D. V. Soa, N. C. Thao,
Eur. Phys. J. C 78, 653 (2018), arXiv:1706.06152 .
\bib{S331} P. V. Dong, N. T. K. Ngan, D. V. Soa, 
 Phys. Rev. D 90, 075019 (2014, arXiv:1407.3839 .
\bib{r331} J. G. Ferreira Jr, P. R. D. Pinheiro, C. A. de S. Pires, and P. S. Rodrigues da Silva, Phys. Rev. D {\bf 84}, 095019 (2011).
\bib{f3311} Renato M. Fonseca, Martin Hirsch, A flipped 331 model, JHEP 08 (2016) 003, arXiv:1606.01109 .
\bib{f3312} T. T. Hong, H. T. Hung, H. H. Phuong, L. T. T. Phuong, L. T. Hue, 
 PTEP 2020 (2020) 4, 043B03, arXiv:2002.06826.
\bib{Dias:2012xp}
A.~G.~Dias, C.~A.~de S.Pires, P.~S.~Rodrigues da Silva and A.~Sampieri,
Phys. Rev. D \textbf{86} (2012), 035007
arXiv:1206.2590.
 \bib{Catano:2012kw}
M.~E.~Catano, R.~Martinez and F.~Ochoa,
Phys. Rev. D \textbf{86} (2012), 073015
arXiv:1206.1966 .
\bib{in331} S. M. Boucenna, J. W. F. Valle, and A. Vicente, Phys. Rev. D
92, 053001 (2015).
\bib{in332} T. P. Nguyen, T. T. Le, T. T. Hong, and L. T. Hue,
 Phys. Rev. D 97 (2018) 7, 073003, arXiv:1802.00429.
\bib{in333} L. T. Hue, H. T. Hung, N. T. Tham, H. N. Long, T.Phong Nguyen,
Phys. Rev. D 104, 033007 (2021), arXiv:2104.01840 . 

\bib{longchang} D. Chang and H. N.  Long,
   Phys. Rev. D {\bf 73}, (2006) 053006,  hep-ph/0603098].
     
\bib{jhep18} A. E. C\'arcamo Hern\'andez, Sergey Kovalenko, H. N. Long, and  Ivan
  Schmidt, 
    J. High Energy Phys. {\bf 07} (2018) 144, arXiv:1705.09169 .

 \bibitem{Casas:2001sr}  	J.~A.~Casas and A.~Ibarra,
   	Nucl. Phys. B \textbf{618} (2001), 171-204
  arXiv:hep-ph/0103065 .

 \bib{Hong28}
  T.~T.~Hong, N.~H.~T.~Nha, T.~P.~Nguyen, L.~T.~T.~Phuong and L.~T.~Hue,
  PTEP \textbf{2022} (2022) no.9, 093B05
  arXiv:2206.08028.
\bib{A41} F. Yin, Phys. Rev. D 75, 073010 (2007). 
 
\bib{A42}  P. V. Dong, L. T. Hue, H. N. Long, D. V. Soa,
 Phys. Rev. D81:053004,2010, arXiv:1001.4625 .

\bib{S31} P. V. Dong, H. N. Long, C. H. Nam and V. V. Vien,
 Phys. Rev. \textbf{D}  85,
(2012) 053001, arXiv:1111.6360. 

\bib{S41} P. V. Dong, H. N. Long, D. V. Soa, V. V. Vien,
Eur. Phys. J. C71:1544,2011, arXiv:1009.2328 .

\bib{D41}V. V. Vien, H. N. Long,
Int. J. Mod. Phys. A 28, No. 32, 1350159 (2013),  arXiv:1312.5034.

\bib{T71} V. V. Vien and H. N. Long, 
  J. High Energy Phys. {\bf 04} (2014) 133,  arXiv:1402.1256.
\bibitem{Froggatt79} C. D. Froggatt and H. B. Nielsen, Nucl. Phys. B 147, 277 (1979).
\bibitem{Leurer93} M. Leurer et. al., Nucl. Phys. B 398, 319 (1993).
\bibitem{Vien22} V. V. Vien, Mod. Phys. Lett. A37 (2022) 2250094.

\bibitem{Longprd19} H. N. Long, N. V. Hop, L. T. Hue, N. H. Thao, A. E. Cárcamo Hernández, 
 Phys. Rev. D 100 (2019) 1, 015004. 
\bibitem{Montero18} J. C. Montero, Ana R. Romero Castellanos, B. L. Sanchez-Vega, Axion dark matter in a 3-3-1 model, Phys. Rev. D 97, 063015 (2018).
\bibitem{Loi21} Duong Van Loi, Cao H. Nam, Phung Van Dong,
 Eur. Phys. J. C 81(2021)591.
 \bibitem{Parker18} R. H. Parker {\it et. al}., Science 360 (2018) 191.
 \bibitem{Isha23} Md. Isha Ali et. al., Eur. Phys. J. C 83 (2023) 60.
 \bibitem{Morel20} L. Morel et.al., Nature 588 (2020) 61.
 \bibitem{Antonio33121} A. E. Cárcamo Hernández, Yocelyne Hidalgo Velásquez, Sergey Kovalenko, H. N. Long, Nicolás A. Pérez-Julve, V. V. Vien, 
 Eur. Phys. J. C 81 (2021) 2, 191, 	arXiv:2002.07347.
 \bibitem{Antonio331d4}A. E. Cárcamo Hernández, H. N. Long, M. L. Mora-Urrutia, N. H. Thao, V. V. Vien, 
 Eur. Phys. J. C 82, No 8 (2022) 769, arXiv:2104.04559.
\bibitem{pq1} R. D. Peccei and H. Quinn, Phys.Rev.Lett. {\bf 38},1440(1977).
	
\bibitem{pq2} R. D. Peccei and H. Quinn, Phys. Rev. D.{\bf 16},1791(1977).

\bibitem{a1}  A. G. Dias, V. Pleitez and M. D. Tonasse, Phys. Rev. D {\bf 67}, 095008 (2003)
arXiv:hep-ph/0211107

\bibitem{a2} A. G. Dias, C. A. de S. Pires and P.~S.~Rodrigues da Silva, 
Phys.\ Rev.\ D {\bf 68} (2003) 115009, arXiv:0309058.
\bibitem{a3} A. G. Dias and V.	Pleitez, Phys. Rev. D {\bf 69} (2004) 077702, arXiv:hep-ph/0308037.
\bibitem{julio1} A. G. Dias, J. Leite, D. D. Lopes, C. C. Nishi, Phys. Rev. D 98, 115017 (2018), arXiv:1810.01893.
\bibitem{julio2} A. G. Dias, J. Leite, José W. F. Valle, C. A. Vaquera-Araujo, Phys. Let.  B 810 (2020) 135829, arXiv:2008.10650.
\bibitem{jpf} J. G. Ferreira, C. A. de S. Pires, J. G. Rodrigues and P.S. Rodrigues da Silva, Phys.\ Lett.\ B {\bf 771} (2017) 199.	
\bibitem{alp331} V. H. Binh, D. T. Binh,  A. E. C\'arcamo Hern\'andez, D. T. Huong, D. V. Soa, and  H. N. Long,  
Phys. Rev. D {\bf 107},  095030 (2023),  arXiv:2007.05004.

\bibitem{giogi} H. Georgi, D. B. Kaplan and L.  Randall,  Phys. Lett. \textbf{B 169} (1986) 73.
\bibitem{jekim} J. E. Kim and G.  Carosi, Rev. Mod. Phys. 82 (2010) 557, Rev. Mod. Phys. 91 (2019) 049902,2019(E), arXiv:0807.3125.

\bibitem{luzio} L. Di Luzio, M.  Giannotti, E. Nardi and L. Visinelli,  Phys. Rept. 870 (2020) 1,  arXiv:: 2003.01100 [hep-ph]
 \bibitem{gu} Pei-Hong Gu, 
JCAP 2016, arXiv:1603.05070.
\bibitem{choi} K. Choi, S. H. Im, H. J. Kim and H. Seong, JHEP 08 (2021) 058,  arXiv: 2106.05816.
 
\bibitem{lh331} H. N. Long, L. T. Hue, Peccei-Quinn mechanism and axion interactions in the 3-3-1 model with
Cosmological Inflation, arXiv:2310.02820 
 
\bibitem{chp2}  S. Scherer, Introduction to Chiral Perturbation Theory,  Adv. Nucl. Phys. 27 (2003) 277, 
arXiv:hep-ph/0210398  

\bibitem{chp3} L.  Di Luzio, G. Martinelli, G.  Piazza, Phys. Rev. Lett. 126, 241801 (2021), arXiv:2101.10330 [hep-ph].
 
 \bibitem{sr} M. Srednicki, Nucl. Phys. B 260 (1985) 689.



\bib{lhc1t}F. Cuypers and S. Davidson,
 Eur. Phys. J. C 2 (1998) 503, [hep-ph/9609487].

\bib{7} J. E. C. Montalvo and M. D. Tonasse, Phys. Rev. D 71
(2005) 095015.

 \bib{lhc1} Le Duc Ninh and H. N. Long,
 Phys. Rev.  D  {\bf 72}, (2005) 075004,  [arXiv: hep-ph/0507069].
 
 \bib{lhc2t} V. Oliveira, C. A. de S. Pires,  Phys. Rev.D 106 (2022) 1, 015031,
	arXiv:2112.03963 [hep-ph]
 

\bib{lhc0} Qing-Hong Cao, Dong-Ming Zhang. 
{\it Collider Phenomenology of the 3-3-1 Model}, arXiv:1611.09337 [hep-ph].

\bib{14} P. H. Frampton, J. T. Liu, B. C. Rasco and D. Ng, Mod.
Phys. Lett. A 9, (1994) 1975.

\bib{lhc2} A. Alves, E. Ramirez Barreto, A. G. Dias,
 Phys. Rev. D 84 (2011) 075013,
 arXiv:1105.4849 .
 		
 \bib{lhc3} E. Ramirez Barreto, Y. A. Coutinho, J. Sá Borges, 
Phys. Rev. D 83:075001,2011, arXiv:1103.1267.

\bib{eec1} H. N. Long and D. V. Soa, 
  Nucl. Phys. {\bf B 601}, (2001) 361, 
 [arXiv:hep-ph/0104150]

\bib{eec2} D. T. Binh, D. T. Huong, Tr. T. Huong, H. N. Long, and D. V. Soa,
 J. Phys. G: Nucl. Part. Phys. {\bf 29}  (2003) 1213, arXiv: hep-ph/0211072.

\bib{eec3} E. Ramirez Barreto, Y. A. Coutinho, J. Sá Borges,
Eur. Phys. J. C 50:909-917,2007, arXiv:hep-ph/0703099.

\bib{neumass1t} F.  Queiroz, C. A. de S. Pires, P. S. Rodrigues da Silva, Phys. Rev. D{\bf 82} (2010), 065018.

\bib{neumass1}   W. Caetano, D. Cogollo, C. A. de S. Pires, P. S. Rodrigues da Silva, Phys. Rev. D 86 (2012),
055021.
\bib{neumassrhn4} N. A. Ky, N. T. H. Van,
Phys. Rev. D72(2005), 115017.

\bib{neumassrhn1}  A. G. Dias, C.A. de S.Pires, P.S. Rodrigues da Silva, Phys. Lett. B628
(2005), 85.
\bib{neumassrhn2} A. G. Dias, C. A.
de S. Pires, P . S. Rodrigues da Silva, Phys. Rev. D82 (2010), 035013.
\bib{neumassrhn3}  D. Cogollo, H. Diniz, C. A. de S.Pires, Phys. Lett. B677 (2009), 338.

\bib{neumassrhn5}  A. Palcu, Mod. Phys. Lett. A21(2006), 2591.

\bib{neumassrhn6}  P. V. Dong, H.
N. Long, Phys. Rev. D77(2008), 057302.

\bib{neumassrhn7}  D. Cogollo, H. Diniz, C. A. de S.Pires, P.S. Rodrigues
da Silva, Eur. Phys. J. C58(2008), 455.


\bib{Pires}  C. A. de S. Pires, {\it Neutrino mass mechanisms in 3-3-1 models: A short review},
  Physics International 2015,
arXiv:1412.1002.



\bib{ISS331}
A. G. Dias, C. A. de S. Pires, P. S. Rodrigues da Silva, A. Sampieri, 
Phys. Rev. D{\bf 86} (2012), 035007.
\bib{moreonISS331}
M. E. Catao, R. Martinez, F. Ochoa, Phys. Rev. D {\bf 86} (2012),073015

\bib{kit1} T. Kitabayashi, M. Yasue, 
Phys.Rev. D63 (2001) 095002,
arXiv:hep-ph/0010087.
\bib{kit2} T. Kitabayashi, M. Yasue,  Phys.Lett. B490 (2000) 236, 
arXiv:hep-ph/0006014
	
\bib{kit3rn} T. Kitabayashi, M. Yasue,
Phys.Lett. B508 (2001) 85,
arXiv:hep-ph/0102228.


\bib{clfv1} L. Lavoura, 
Eur.Phys.J.C29:191-195,2003, 
	arXiv:hep-ph/0302221.
	
\bib{clfv2} L. T. Hue, L. D. Ninh, T. T. Thuc, N. T. T. Dat,
 Eur. Phys. J. C (2018) 78:128,  arXiv:1708.09723 [hep-ph]. 

\bib{clfv3} L. T. Hue, H. T. Hung, N. T. Tham, H. N. Long, T.Phong Nguyen,
Phys. Rev. D 104, 033007 (2021), arXiv:2104.01840.

\bib{amm1}  N. A.  Ky,  H. N. Long, and  D. V. Soa,
  Phys. Lett. {\bf B 486}, (2000) 140, arXiv: hep-ph/0007010.


\bib{amm2} D. P. Aguillard et al. Muon $g-2$, Phys. Rev. Lett. 131 (2023) no.16, 161802
[arXiv:2308.06230 [hep-ex]].

\bib{amm3} D. P. Aguillard et al, Measurement of the Positive Muon AMM to 0.46 ppm, 
Phys. Rev. Lett. 131 (2023) 161802.

\bib{amm4} T.  Aoyama, T. Kinoshita and M. Nio, Theory of the AMM of the Electron, Atoms 7, 28

\bib{amm5} T. T. Hong, L. T. T. Phuong, T.Phong Nguyen, N. H. T. Nha, L. T. Hue,
arXiv:2404.05524 [hep-ph].


\bib{amm6} C. Kelso, H. N. Long, R. Martinez, Farinaldo S. Queiroz, 
Phys. Rev. D 90, 113011 (2014), arXiv:1408.6203.



\bib{amm7} A. E. Cárcamo Hernández, D. T. Huong, H. N. Long, 
Phys. Rev. D 102, 055002 (2020), arXiv:1910.12877.

\bibitem{amm8} A. E. Cárcamo Hernández, Yocelyne Hidalgo Velásquez, Sergey Kovalenko, H. N. Long, Nicolás A. Pérez-Julve, V. V. Vien, 
Eur. Phys. J. C 81, (2021) 191, 
arXiv:2002.07347.


\bib{dm1} D. Fregolente and M. D. Tonasse, Phys. Lett. B555 (2003) 7.

\bib{dm2} H. N. Long and Nguyen Quynh Lan,
   Europhys. Lett. {\bf 64} (2003) 571,  arXiv: hep-ph/0309038.
   

\bib{dm3} P. V. Dong, N. T. K. Ngan, D. V. Soa, 
Phys. Rev. D 90, 075019 (2014),  arXiv:1407.3839.

 	
\bib{inf1} P. V. Dong, D. T. Huong, D. A. Camargo, Farinaldo S. Queiroz, José W. F. Valle,
Phys. Rev. D 99, 055040 (2019), arXiv:1805.08251.

\bib{inf2} D. T. Huong, P. V. Dong, C. S. Kim, N. T. Thuy,
Phys. Rev. D 91, 055023 (2015), arXiv:1501.00543.

\bib{cp1} V. Q. Phong, H. N. Long and Vo Thanh Van, 
Phys. Rev.
\textbf{D 88}, 096009 (2013), [arXiv:1309.0355.

\bib{cp2} V. Q. Phong, H. N. Long, Vo Thanh Van,  Le Hoang
Minh,
  Eur. Phys. J.
\textbf{C 75},  (2015) 342, arXiv:1409.0750.


\bib{cp3} V. Q. Phong, , N. T. Tuong, N. C. Thao, H. N. Long,
 Phys. Rev. D 99, 015035 (2019), arXiv:1805.09610.


\bib{lepg1} D. T. Huong, H. N. Long, 
J. Phys. G 38:015202,2011, arXiv:1004.1246.


\bib{lepg2} A. E. Cárcamo Hernández, D. T. Huong, H. N. Long,
Phys. Rev. D 102, 055002 (2020), arXiv:1910.12877.


\bib{bau1} V. Q. Phong, H. N. Long, V. T. Van, N. C. Thanh,
Phys.Rev.D 90 (2014) 8, 085019 ,
arXiv:1408.5657.



\end{thebibliography}
\end{document}